\documentclass[12pt]{article}

\usepackage{amsfonts}
\usepackage{amssymb}
\usepackage{amsmath}
\usepackage{latexsym}

\usepackage{graphicx}

\newcommand{\al}{\alpha}

\newcommand{\pa}{\partial}

\newcommand{\om}{\omega}
\newcommand{\de}{\delta}

\newcommand{\rar}{\rightarrow}

\allowdisplaybreaks
\newcounter{mycount}

\begin{document}

\begin{titlepage}
\vspace{-1mm}
\begin{flushright}
Preprint ICN-UNAM 04-06\\
June 22, 2004
\end{flushright}
\vspace{1mm}
\begin{center}
 {\bf\Large\sf The conformally invariant Laplace-Beltrami operator and
 factor ordering}
\end{center}

\begin{center}
{\bf Michael P. Ryan, Jr.}{\normalsize
\footnote{ryan@nuclecu.unam.mx}}
 and
 {\bf Alexander V.~Turbiner{\normalsize \footnote{
turbiner@nuclecu.unam.mx}${}^{,} $\footnote{On leave of absence
from the Institute for Theoretical and Experimental Physics, \\
\indent \hspace{5pt} Moscow 117259, Russia.}}}
\\[2mm]
{\em Instituto de Ciencias Nucleares, UNAM, A.P. 70-543,\\ 04510
Mexico D.F., Mexico} \\ \vspace{2mm}

{\bf\large Abstract}
\end{center}

\begin{quote}
In quantum mechanics the kinetic energy term for a single particle
is usually written in the form of the Laplace-Beltrami operator.
This operator is a factor ordering of the classical kinetic
energy.  We investigate other relatively simple factor orderings
and show that the {\it only\/} other solution for a conformally
flat metric is the conformally invariant Laplace-Beltrami
operator. For non-conformally-flat metrics this type of factor
ordering fails, by just one term, to give the conformally
invariant Laplace-Beltrami operator.
\end{quote}
\vskip 1.5cm

\centerline{\it Phys. Lett. A (in print)}

\vskip .5cm

\end{titlepage}

Factor ordering is a complication that has bedeviled quantum
mechanics since its inception.  The modern literature on this
subject, while considerable, seems, outside of elementary texts,
to be mostly concentrated in the field of canonical quantum
gravity, where many factor orderings of the momentum term of the
Wheeler-DeWitt equation have been proposed. These proposals go
back to DeWitt \cite{dewit}, and we will mention only a few
articles that are relevant to the problem in ordinary quantum
mechanics.  Komar \cite{komar}, basing his arguments on Pauli's
work on quantum mechanics \cite{pauli} suggests that the factor
ordering of momentum terms must be equivalent to using momentum
vectors lying along the Killing vectors of the configuration space
(this assumes that the space admits a Riemannian metric). Many
factor orderings of the Wheeler-DeWitt momentum terms exist,
several for quantum cosmology, two important proposals being those
of Misner \cite{misner} and Hartle and Hawking \cite{hh}. The main point
is that it is never entirely clear when one passes from some
product of classical quantities to the same product of their
quantum analogues, exactly which ordering of the product to take,
and there is an even more difficult problem, whether to
interpolate functions of coordinate operators between momentum
operators in such a way that the classical limit is preserved.
This last point can be a dangerous procedure. This type of factor
ordering can be used to transform the Hamiltonian of a free
particle into that of a harmonic oscillator \cite{karel},
generating a discrete spectrum from a continuous one.  That is, if
one writes the quantum analogue of a one-dimensional free particle
Hamiltonian, $H = p_x^2/2m$, one can choose either
\begin{equation}
\label {freep}
 \hat H \psi = -\frac{1}{2m} \partial_x^2 \psi\ ,
\end{equation}
or
\begin{equation}
\label {newfre}
 \hat H \psi = -\frac{1}{2m} \frac{1}{f(x) h(x)} \pa_x [f\partial_x
 (h\psi)]\ ,
\end{equation}
where $f$ and $h$ are arbitrary functions. If we now choose $f = 1/h^2$
and $h = \exp (-\frac{m\om}{2}x^2)$, we have
\begin{equation}
\label{harm}
 \hat H = -\frac{\hbar^2}{2m} \partial_x^2 +
 \frac{1}{2} m\om^2 x^2 + \frac{\omega}{2}\ ,
\end{equation}
a harmonic oscillator with a slightly shifted spectrum.

There are, of course, many possible factor orderings of the
original free particle Hamiltonian, including, in general, any
algebraic combination of the commutator $[\hat x, \hat p_x]$
(although one usually prefers to keep the Hamiltonian as a second
order differential operator). The usual solution to this problem
is to exploit experimental evidence to exclude all but one (or a
very reduced set) of the possible factor orderings.

One of the most common factor ordering problems is not usually
regarded as being related to factor ordering.  In a space of
dimension $n$ parametrized by non-Cartesian coordinates (which may
or may not be flat), the usual classical expression for the
kinetic energy term of a one-particle Hamiltonian for a phase
space $(p_a, q^a)$ is
\begin{equation}
\label {claham}
 \frac {1}{2m} g^{ab}p_a p_b. \qquad a,b = 1, \ldots, n\ ,
\end{equation}
$g_{ab} = g_{ab}(q^c)$. This may be converted to an
operator by taking $\hat p_a = -i\pa /
\pa q^a$. The simplest factor ordering, $g^{ab}\hat p_a \hat p^b$
does not seem to give the correct Hamiltonian in most cases. For
example, for spherical coordinates in a flat, three-dimensional
space, this factor ordering gives
\begin{equation}
-\frac{1}{2m} \left [ \frac{\partial^2}{\partial r^2} +
\frac{1}{r^2} \frac{\partial^2}{\partial \theta^2} + \frac{1}{r^2
\sin^2 \theta} \frac{\partial^2}{\partial \varphi^2} \right ]\ ,
\end{equation}
and this expression, used in the Hamiltonian of a hydrogen atom, gives an
energy spectrum that, for the magnetic quantum number, $m$, equal to zero is
\begin{equation}
E_{nlm} = \frac{me^4}{2[n - l - 1/2 + \sqrt{l^2 + 1/4}]^2},
\end{equation}
and for $|m| > 0$ is
\begin{equation}
E_{nlm} = \frac{me^4}{2[n - l - 1/2 + \sqrt{\{l + \frac{1}{2}(1 +
\sqrt{4m^2 + 1})\}^2 + 1/4}]^2},
\end{equation}
with no restriction that $|m|$ be less than $l$.  These
expressions are only in accord with experiment for ($m = 0$)
s-states. The wave functions are similar to the usual ones, but
for $m = 0$ the Legendre polynomials are replaced by Chebyshev
polynomials and the radial functions have the form
$e^{-r/a_0}r^{(1 + 2\sqrt{l^2 + 1/4})/2} L_{n - l -
1}^{(2\sqrt{l^2 + 1/4})}(r/a_0)$, $a_0$ the Bohr radius and
$L^{(\alpha)}_n$ Laguerre polynomials. For $m \neq 0$ the angular
functions are based on $e^{im\varphi}(\sin \theta)^{(1 + \sqrt{1 +
4m^2})/4} G^{(1 + \sqrt{1 + 4m^2})/2}_l (\cos \theta)$,
$G^{\sigma}_{\lambda}$ Gegenbauer polynomials and radial functions
that are proportional to
\begin{equation}
e^{-r/a_0}r^{(1 + \sqrt{(1/2 + \sqrt{[l + \frac{1}{2}(1 + \sqrt{1
+ 4m^2}})]^2 + 1/4}} L^{(2\sqrt{[l + \frac{1}{2}(1 + \sqrt{1 +
4m^2}]^2 + 1/4})}_{n - l - 1}\ .
\end{equation}
The usual solution to this problem is to write
the operator form of (\ref{claham}) in terms of a Laplace-Beltrami operator,
\begin{equation}
\frac{1}{2m} \frac{1}{\sqrt{g}} \hat p_a (\sqrt{g} g^{ab})\hat p_b,
\label {lapbel}
\end{equation}
which is clearly a factor ordering of (\ref{claham}). This factor
ordering, which is invariant under changes of the configuration
variables $q^c$, has been very successful in constructing
Hamiltonians that are in accord with experiment in many cases.

For a space that is not flat there is another coordinate-invariant
operator besides the Laplace-Beltrami operator that is sometimes
used as the quantum version of $g^{ab}p_a p_b$, the conformally
invariant Laplace-Beltrami operator,
\begin{equation}
\label {conflb}
 \frac{1}{\sqrt{g}} \hat p_a (\sqrt{g} g^{ab} \hat
 p_b)\ - \ \frac{n - 2}{4(n - 1)}R\ ,
\end{equation}
(for $n > 1$), where $R$ is the Ricci scalar or scalar curvature.
This curvature, as
a function of the metric $g_{ab}$ only, is
\begin{equation}
\label{r}
 R = -g^{ab} \frac{g_{,a,b}}{g} + \frac{3}{4} g^{ab}
\frac{g_{,a} g_{,b}}{g^2} -\frac{g_{,b}}{g} g^{ab}_{\,\,\,\, ,a}
-\frac{1}{2} g^{ab}_{\,\,\,\, ,l}g_{ra,b} g^{lr} +\frac{1}{4}
g^{ab}_{\,\,\,\, ,l} g_{ab,r} g^{lr}\ .
\end{equation}

It is interesting to note that over the years there has been some
interest in relating the operator (\ref{conflb}) to time
reparametrization invariance.  In a series of articles going back
to Misner \cite{misner}, (see Refs. \cite{mossy} and \cite
{hali}), a possible action for a relativistic particle moving in a
curved background (and the similar cosmological minisuperspace
actions) have a Hamiltonian constraint multiplied by a ``lapse
function" $N$ that serves as a Lagrange multiplier.  If one
rescales $N$ by an arbitrary function of the coordinates and
insists that the quantum theory generated by the action be
invariant under this rescaling, the momentum part of the quantum
Hamiltonian constraint is {\it necessarily\/} (\ref{conflb}),
where $n$ is either the dimension of the space in which the
particle moves or the dimension of the minisuperspace.

Since the conformally invariant Laplace-Beltrami operator has a
similar relationship to the ordinary Laplace-Beltrami operator
(\ref{lapbel}) as that between (\ref{freep}) and (\ref{newfre}),
we can ask what operators can be constructed from (\ref{claham})
that are similar to that used to construct (\ref{newfre}) and
whether (\ref{conflb}) is included in this set of operators, that
is, investigating factor orderings of the type
\begin{equation}
g^{ab}p_a p_b \rar \frac{1}{H(g_{cd})} \pa_a [A(g_{cd}) \pa_b
B(g_{cd})]\ ,
\end{equation}
where $H(g_{cd})$, $A(g_{cd})$, $B(g_{cd})$ are arbitrary
functions of the metric with the constraint $AB = g^{ab}H$.  The
simplest functions that we can use are related to powers of the
determinant $g$ of $g_{ab}$. One possibility is
\begin{equation}
\label{galbe}
 g^{ab} p_a p_b \rar -\frac{1}{\sqrt{g}
g^{\tilde \al + \tilde \beta}} \pa_a [g^{\tilde \al} \sqrt{g}
g^{ab} \pa_b g^{\tilde \beta}]\ .
\end{equation}
In the general case no choice of $\tilde \alpha$ and $\tilde \beta$ can give
an expression that includes (\ref{conflb}), but if we consider only
conformally flat metrics, $g_{ab} = \phi^2 \delta_{ab}$, we find that
\begin{equation}
{\rm det} \, g_{ab} = \phi^{2n} = g\ ,
\end{equation}
so $g_{ab} = g^{1/n} \de_{ab}$, $g^{ab} = g^{-1/n}\de_{ab}$, and
the scalar curvature is
\begin{equation}
\label{confr}
 R = -\left ( 1 - \frac{1}{n}\right ) g^{-1/n} \left
[\frac{\pa^2 g}{g} - \left (\frac{1}{2n} - \frac{3}{4}\right )
\frac{(\pa g)^2}{g^2} \right ]\ ,
\end{equation}
where $\pa^2 g/g \equiv (g_{,a,b}/g)\de_{ab}$, $(\pa g)^2/g^2
\equiv (g_{,a} g_{,b}/g^2)\de_{ab}$.

Assuming we want this factor ordering of $g^{ab} p_a p_b$ to give
\begin{equation}
\label{newlbr}
 \frac{1}{\sqrt{g}} \pa_a [\sqrt{g} g^{ab}\pa_b] + CR\ ,
\end{equation}
$C =$ constant. Then we must eliminate first-order derivative
terms, which implies $\tilde \al + 2\tilde \beta = 0$.  We find
that we have (\ref{newlbr}) with $C = -\tilde \beta (1 - 1/n)$ and
\begin{equation}
-\tilde \beta  \left( 1 = \frac{1}{n} \right )^2 \left (
\frac{1}{2n} + \frac{3}{4} \right ) = \tilde \beta \left (\tilde
\al + \frac{1}{2} \right ) + \tilde \beta (\tilde \beta - 1) -
\frac{\tilde \beta}{n}\ ,
\end{equation}
or
\begin{equation}
\tilde \beta = -\frac{1}{2n} + \frac{1}{4}, \qquad \tilde \al =
\frac{1}{n} - \frac{1}{2}\ ,
\end{equation}
and finally,
\begin{equation}
C = -\frac{1}{4} \frac{n - 2}{n - 1}\ .
\end{equation}
There is only one other solution, $\tilde \al = \tilde \beta = C =
0$. This means that there are only two possible solutions leading
to (\ref{newlbr}), the ordinary Laplace-Beltrami operator ($C =
0$) or the conformally invariant operator.

For a general metric $g_{ab}$ there are many more possible factor
orderings. We can attempt a factor ordering that makes use of
powers of the metric components, where we can use matrix
identities such as ($\delta^i_j$ the Kronecker delta and
$\varepsilon^{ij \cdots k}$ the totally antisymmetric Levi-Civita
symbol)
\begin{equation}
\delta^r_a \delta^s_r \cdots \delta^k_n = \underbrace {g_{aq}
g_{cd} \cdots g_{fy}}_{m_1 + m_2} \underbrace {g^{qa} g^{de}
\cdots g^{sv}}_{m_1} \underbrace {g^{wr} g^{tn} \cdots
g^{yk}}_{m_2}\ ,
\end{equation}
\begin{equation}
\frac{1}{n! g} \varepsilon^{ij \cdots k} \varepsilon^{l_1 m_1
\cdots n_1} \underbrace {g_{il_1} \cdots g_{p_1 q_1}}_{k_1}
\underbrace {g_{w_1 v} \cdots g_{k n_1}}_{k_2} = 1\ ,
\end{equation}
\begin{equation}
\underbrace {g_{ab} g^{lm} \cdots g_{es}}_{q_1} \underbrace{g_{vw}
g_{rt} \cdots g_{fg}}_{q_2} \underbrace {g^{aq} g^{lm} \cdots
g^{es} g^{vw} \cdots g^{fg}}_{q_1 + q_2} = n^{q_1 + q_2}\ .
\end{equation}
The most general factor ordering using these expressions we have
been able to construct is
\[
g^{ab} p_a p_b \rar \frac{(g_{lr})^{q_1 + q_2} (g_{ac})^{r_1 +
r_2} (g_{bq})^{m_1 + m_2} \varepsilon^{ij \cdots k}
\varepsilon^{l_1 m_1 \cdots n_1}}{n^{q_1 + q_2} g^{\al + \beta +
1/2} n! g}
\]
\[
\times \pa_d [g^{ab} g^{\alpha + 1/2} (g^{lr})^{q_1}
(g^{st})^{r_1})(g_{qv})^{m_1} \underbrace{g_{il_1} \cdots
g_{pq_1}}_{k_1}
\]
\begin{equation}
 \times \pa_c (g^{wc})^{m_2}(g^{lr})^{q_2} (g^{yd})^{r_2}
\underbrace{g_{wv_1} \cdots g_{kn_1}}_{k_2} g^{\beta}]\ , \label{genlb}
\end{equation}
where, for example, $(g_{bq})^{m_1 + m_2}$ is shorthand for
$\underbrace{g_{bq} g_{cd} \cdots g_{fg}}_{m_1 + m_2}$.

Considerable algebra leads to two conditions on the nine
constants, $q_1$, $q_2$, $r_1$, $r_2$, $m_1$, $m_2$, $\alpha$,
$\beta$, $k_1$ ($k_2 = n - k_1$), that force the terms linear in
derivatives to vanish.  There are six independent terms in $R$ as
given in (\ref{r}), and our factor ordering gives a ``potential''
term (which we would like to be $CR$, $C$ the constant in [\ref
{newlbr}]) with seven independent terms, each multiplied by
combinations of the seven independent constants that remain after
the two conditions from putting the terms linear in derivatives to
zero are satisfied.  It is remarkable that six of these seven terms
have {\it exactly\/} the form of the six terms in $R$ and that there
is only a single extra term that has the form $g^{ab}_{\,\,\,\, ,b} g_{ac}
g^{cl}_{\,\,\,\, ,l}$. If it were not for
this seventh term, we could find a set of constants that would give $CR$.
Unfortunately, the seven equations that
relate the seven constants to each other and to $C$ are
inconsistent except for $C = 0$, so the only possible solution is
the ordinary Laplace-Beltrami operator.

If we explicitly reduce the metric to conformally flat form, we
have the two solutions given above, so the factor ordering given
in (\ref{genlb}) is consistent with our previous result.  Since a
conformally flat metric is characterized in a coordinate invariant
way by a zero Weyl tensor, investigating the relation between the
conformally-flat calculation and the general calculation in terms
of the Weyl tensor may give some insights into the reasons the
factor ordering we chose does not give the conformally invariant
Laplace-Beltrami operator.

Finally, we have chosen a specific factor ordering similar to that
given in (\ref{newfre}), where we have taken powers of the metric
components replacing the $f$ and $h$ functions of
Eq.(\ref{newfre}).  Of course, this is a very specific factor
ordering, and there is no reason to believe that there is {\it
no\/} factor ordering that can give the conformally invariant
Laplace-Beltrami operator. In fact, Moss and Shiiki \cite{ms} have
used the possibility of terms proportional to the commutator $[\hat
p_c, \hat q_d] = -i\delta_{cd}$ to write a factor ordering that
gives the conformally invariant Laplace-Beltrami operator by
adding the term $R^a_{\, \, b} [\hat p_a, \hat q_b]$ to the ordinary
Laplace-Beltrami operator. A search for a more general factor
ordering could be the subject of future research.

\section*{Acknowledgments}

M.R. thanks K. Kucha\v r for interesting discussions. We also
thank FENOMEC, UNAM for financial support. One of us (A.V.T.) is
supported in part by CONACyT grant {\it 25427-E}.


\begin{thebibliography}{99}

\bibitem{dewit}
      B.~DeWitt, {\it Phys. Rev.\ \bf 160}, 1113 (1967).

\bibitem{komar}
         A.~Komar, {\it Phys. Rev.\ \bf 170}, 1195 (1968).

\bibitem{pauli}
         W.~Pauli, in {\it Handbook der Physik \bf 5/1}, p.40, S.
         Fl\" ugge, Ed. (Springer Verlag, Berlin, 1958).

\bibitem{misner}
         C.~Misner, in {\it Magic without Magic: John Archibald
         Wheeler}, J. Klauder, Ed. (Freeman, San Francisco, 1972).

\bibitem{hh}
         J.~Hartle and S.~Hawking, {\it Phys. Rev. D\ \bf 28}, 2960 (1983).

\bibitem{karel} See K. Kucha\v r, Lecture notes on quantum theory, University
of Utah (unpublished).

\bibitem{mossy} I. G.~Moss, {\it Ann. Inst. Henri Poincar\' e\ \bf 49}, 341 (1988).

\bibitem{hali} J. J.~Halliwell, {\it Phys. Rev. D\ \bf 38}, 2468 (1988).

\bibitem{ms} I.G.~Moss, and N.~Shiiki, {\it Nucl.Phys.\ \bf B565}, 345 (2000).







\end{thebibliography}
\end{document}